\documentclass[12pt]{article}

\usepackage[margin=1in]{geometry}

\usepackage[nottoc]{tocbibind}

\usepackage[utf8]{inputenc}       
\usepackage[T1]{fontenc}          

\usepackage{amsmath,amssymb,amsfonts,amsthm}  
\usepackage{mathtools}                         
\usepackage{bm}                                
\usepackage{dsfont}                            

\usepackage[round]{natbib}       

\usepackage[ruled,vlined]{algorithm2e}

\usepackage{graphicx}

\usepackage{caption}
\usepackage{hyperref}  

\usepackage{booktabs}             

\usepackage{xcolor}               
\usepackage[normalem]{ulem}       

\usepackage{enumerate}            

\newcommand{\Rset}{\mathbb{R}}

\newcommand{\Sset}{\mathbb{S}}

\DeclareMathOperator{\E}{E}

\newcommand{\normx}[1]{\Vert#1\Vert}

\newcommand{\Tau}{\mathcal{T}}
\newcommand{\noop}[1]{}           

\theoremstyle{plain}

\theoremstyle{definition}

\theoremstyle{remark}

\newcommand{\blind}{1}

\begin{document}

\if1\blind
{%
  \title{\bf Club Exco: \\ clustering brain extreme communities from multi-channel EEG data}
  \author{Matheus B. Guerrero$^{1}$, Paolo V. Redondo$^{2}$,\\[2.5pt] Marco A. Pinto-Orellana$^{3}$, Beth A. Lopour$^{3}$,\\[2.5pt] Hernando Ombao$^{2}$, and Raphaël Huser$^{2}$ \\[5pt]
    {\small{$^{1}$Department of Mathematics, NSM, CSUF}} \\[1pt]
    {\small{$^{2}$Statistics Program, CEMSE Division, KAUST}}\\[1pt]
    {\small{$^{3}$Department of Biomedical Engineering, UCI}}}
  \maketitle
} \fi

\if0\blind
{%
  \bigskip
  \begin{center}
    {\LARGE\bf Club Exco: clustering brain extreme communities from multi-channel EEG data}
  \end{center}
  \medskip
} \fi

\begin{abstract}
\noindent Current methods for clustering brain networks over time often rely on cross-dependence measures computed from the entire range of EEG signals, which can obscure information specific to extreme neural activity. To overcome this, we introduce Club Exco, a novel clustering method grounded in extreme value theory, designed to detect brain communities with co-occurring high-amplitude EEG events. By focusing on tail behavior, Club Exco isolates extreme-value synchrony across channels, offering new insights into seizure dynamics. We apply Club Exco to neonatal EEG recordings from 30 patients (13 seizure-free and 17 with clinically confirmed seizures). Our method identifies robust ``brain extreme communities'' and constructs Extreme Connectivity Persistence matrices that summarize how often channels exhibit synchronous extremes across time. Seizure patients exhibit more persistent and variable clustering among non-adjacent regions, suggesting seizure propagation, while non-seizure patients show more consistent clustering in anatomically adjacent regions. Compared to coherence-based methods (e.g., Hierarchical Cluster Coherence procedure), Club Exco captures distinct, seizure-associated connectivity patterns, especially in high-amplitude segments. These results highlight Club Exco's potential to characterize extreme neural events and inform clinical understanding of seizure localization and spread.
\end{abstract}


\clearpage
\newpage
\section{Introduction\label{chap:intro}}
Epilepsy is a major neurological disorder characterized by recurrent unprovoked seizures, with neonatal seizures representing a critical subset requiring specialized analysis \citep{karamian2024}. Neonatal seizures, often subtle and electrographic, are the most common neurological emergency in newborns, affecting 1 to 5 per 1,000 live births \citep{Saral2023NeonatalSeizure, Patrizi2003NeonatalSeizures}. Unlike adult seizures, neonatal events often lack visible clinical correlates, with 70–90\% exhibiting no observable behavioral manifestations \citep{Shellhaas2007NeonatalSeizures}. Therefore, electroencephalograms (EEGs), which record electrical brain activity noninvasively through scalp-placed electrodes, become essential tools for detecting neonatal seizures \citep{Ramantani2022NeonatalSeizures, Nagarajan2025StatusEpilepticus}. EEG signals reflect attenuated and distorted cortical neuronal activity that reaches the scalp, providing real-time insights into seizures with high temporal resolution—far superior to functional neuroimaging modalities such as functional magnetic resonance imaging (fMRI). However, analyzing EEG data presents significant challenges due to their highly nonstationary and potentially nonlinear nature. Furthermore, EEG recordings commonly contain artifacts, signals unrelated to brain activity, such as eye movements, muscle activities, and electromagnetic interference, that require sophisticated preprocessing techniques \citep{Kaya21}. Thus, effective EEG analysis relies heavily on preprocessing pipelines and advanced statistical methods to reliably detect and interpret seizure activity \citep{Ombao_book}, especially for neonates.

In particular, changes in the brain signals due to seizures are more appropriately modeled as dependence networks \citep{ModelFocalSeizure-Liou-2020}. Links in such networks have different patterns with varying degrees of strength depending on the condition, as \cite{TemporalDynamicsF-Smith-2020} found in pediatric subjects during sleep or awake states. Furthermore, along the classical epileptogenic zone hypothesis that suggests that a particular tissue is responsible for causing seizures and upon its rejection, the subject would not have further seizure events \citep{PresurgicalEvaluationEpilepsy-Rosenow-2001}, network models also indicate that the coordinated activity among neighbor tissue also has a predominant role in seizure generation and can be a valuable biomarker \citep{GraphTheoreticalFast-Weiss-2022, InterictalStereotacticEEG-Lagarde-2018, EarlySeizureSpread-Andrews-2020}. For instance, \cite{IctalConnectivityChildhood-Tenney-2018} showed a relationship between connectivity patterns and treatment effectiveness: patients with less effective treatment response denoted lower parietal connectivity at low frequencies and high frontal connectivity at high frequencies compared to control groups.

Nevertheless, there is no consensus in the approach to quantify such inter-channel dependence networks \citep{SeizureOnsetZone-Balaji-2022}, as cross-frequency coupling \citep{EpileptogenicZoneLocation-Liu-2021}, frequency-filtered Granger causality \citep{TimeVaryingConnectivity-Ghumare-2018,HighFreqOsc-Li-2022}, partial directed coherence \citep{TimeVaryingConnectivity-Ghumare-2018}, average coherence across neighbor channels \citep{NormativeIntracranialEEG-Bernabei-2022}, and cross-correlation \citep{StrengthStabilityEEG-Shrey-2018} have been used in the literature.

A persistent challenge in EEG analysis lies in the statistical modeling of EEG signal distributions, particularly in capturing their non-Gaussian and heavy-tailed characteristics. As mentioned before, traditional methods, including cross-correlation, coherence, and Gaussian graphical models \citep{Ombao2024SpectralDependence, zhang20ggm}, focus on central tendencies, neglecting neuronal oscillations' heavy-tailed and non-Gaussian nature. This oversight is particularly consequential in neonates, where burst-suppression patterns—characterized by alternating periods of high-amplitude activity (bursts) and near-flatline EEG (suppression)—and transient extreme amplitudes are hallmarks of hypoxic-ischemic encephalopathy, which is a type of brain injury caused by insufficient oxygen and blood flow to the brain around the time of birth \citep{Shellhaas2007NeonatalSeizures, Kharoshankaya2016}. It is worth mentioning that a more general method that does not necessarily rely on a Gaussian setup and might be adapted to capture tail patterns is the Short-Time Direct Directed Transfer Function (SdDTF) \citep{korzeniewska2007dynamics}. However, due to its model and especially computational complexity, it is outside the current work's scope.

Addressing the limitations of traditional EEG analysis, we propose the ``Clustering of Brain Extreme Communities'' (Club Exco) method, grounded in extreme value theory (EVT). EVT—traditionally applied in environmental sciences \citep{davison15,davisonhuser19}, finance \citep{daouia114,gong22}, and public health \citep{Thomas2016EVT,Thomas2022InfluenzaEVT}—provides a unified framework for modeling rare, extreme events in complex data. \cite{guerrero23} and \cite{redondo2024statistics} pioneered early EEG applications of EVT by using a conditional extreme-value model to capture the extremal dependence structure before and after seizures in a human patient. More recently, \cite{boulin2025high} adapted an Asymptotic Independent–Block clustering model to high-dimensional settings, including an illustrative EEG case study. In contrast, Club Exco leverages peaks-over-threshold methods to capture tail dependencies directly in continuous time series, avoiding block-maxima aggregation. Club Exco utilizes EVT principles through spherical $k$-means clustering to identify EEG channel groups exhibiting synchronized extreme amplitudes, referred to as ``brain extreme communities.'' Unlike conventional clustering methods based on central tendencies and spectral coherence, such as the Hierarchical Cluster Coherence (HCC) method \citep{euanHCC}, Club Exco explicitly captures extreme dependencies that traditional approaches overlook. We apply Club Exco in a sliding-window framework to investigate dynamical changes governed by extreme behavior during neonatal seizure episodes. Moreover, Club Exco addresses key clinical objectives related to seizure localization: it identifies channels exhibiting synchronous extreme behavior, detects concomitant extreme amplitudes across channels, and evaluates whether clustering based on extreme communities can enhance machine-learning algorithms for pinpointing seizure foci. These objectives culminate in a dimensionality reduction strategy that uncovers sparse, extremal connectivity structures—patterns we investigate empirically in our neonatal EEG data analysis.

Applying Club Exco to neonatal EEG data, we uncovered patterns in how brain regions co-activate at extreme signal amplitudes during seizure and non-seizure states. In seizure patients, clusters of channels exhibited persistent and variable co-occurrence of high-amplitude activity—especially across non-adjacent regions—suggesting dynamic seizure propagation. In contrast, non-seizure patients displayed more consistent and anatomically localized extreme connectivity. We quantified these patterns using the Extreme Connectivity Persistence (ECP) matrix, which revealed group-level and patient-specific signatures. Compared to HCC, a coherence-based clustering approach, Club Exco captured seizure-associated connectivity differences—particularly in high-amplitude regimes—that the HCC method did not, underscoring the utility of extremal dependence structures for uncovering clinically relevant neural dynamics.

The remainder of this article is structured as follows. In Section~\ref{sec:methodology}, we briefly describe the multivariate extreme value theory and our proposed clustering procedure, Club Exco, for multi-channel EEG data. In Section~\ref{sec:application}, we use the Club Exco method to address questions about the brain network of neonatal patients with seizures. We also compare the results of Club Exco with those from the classical HCC method. We provide our conclusions in Section~\ref{sec:conclusion}. Further details on the data analysis are contained in the Supplementary Material.

\section{Clustering brain extreme communities\label{sec:methodology}}
This section details the general setting and provides the fundamentals of multivariate extreme value theory. It also introduces our proposed clustering approach used to identify concomitant extremes in multi-channel EEG signals. For recent and comprehensive reviews on EVT, we refer to \cite{davison15} and \cite{engelkereview}, while for a study on existing methods for concomitant extremes, see \cite{sabourinthesis}.

\subsection{General data setting}
Let $\bm{X}^{t} = \left(X_1^t, \ldots, X_D^t\right)^{\intercal}$ be the $D$--dimensional vector of amplitudes (i.e., absolute values of a zero-mean time series: $X_{d}^{t} = \vert U_{d}^{t} - \bar{U}_d \vert$, where $U_{d}$ is the original EEG signal and $\bar{U}_d$ is its mean; $t = 1,\ldots, T$, $d = 1, \ldots, D$) from a multi-channel EEG recordings on $D$ channels at time $t$, with $t = 1, \ldots, T$. We focus on absolute signal amplitudes rather than distinguishing large positive from large negative deviations, since EEGs are oscillatory and often modeled as zero-mean stochastic processes. This approach emphasizes the upper-tail behavior—i.e., large amplitudes—while avoiding the need to model the mean or asymmetries between tails. Because EEG signals are measured relative to a chosen reference electrode (which can flip polarity), using absolute values provides consistency across different reference choices and channels, ensuring that large deflections are treated equivalently. Moreover, since extremes are inherently rare, combining information from both tails enhances the detection of meaningful patterns in brain-network activity.

Recall that EEG signals are, in general, nonstationary. Therefore, we consider only short stretches of the time series, e.g., $10$-second sliding windows, where the EEG is quasi-stationary. In the following definitions and theoretical results, we assume that $\{\bm{X}^t\}_{t=1}^{T}$ is a multivariate stationary time series.

\begin{figure}
\centering
\includegraphics[width=\textwidth]{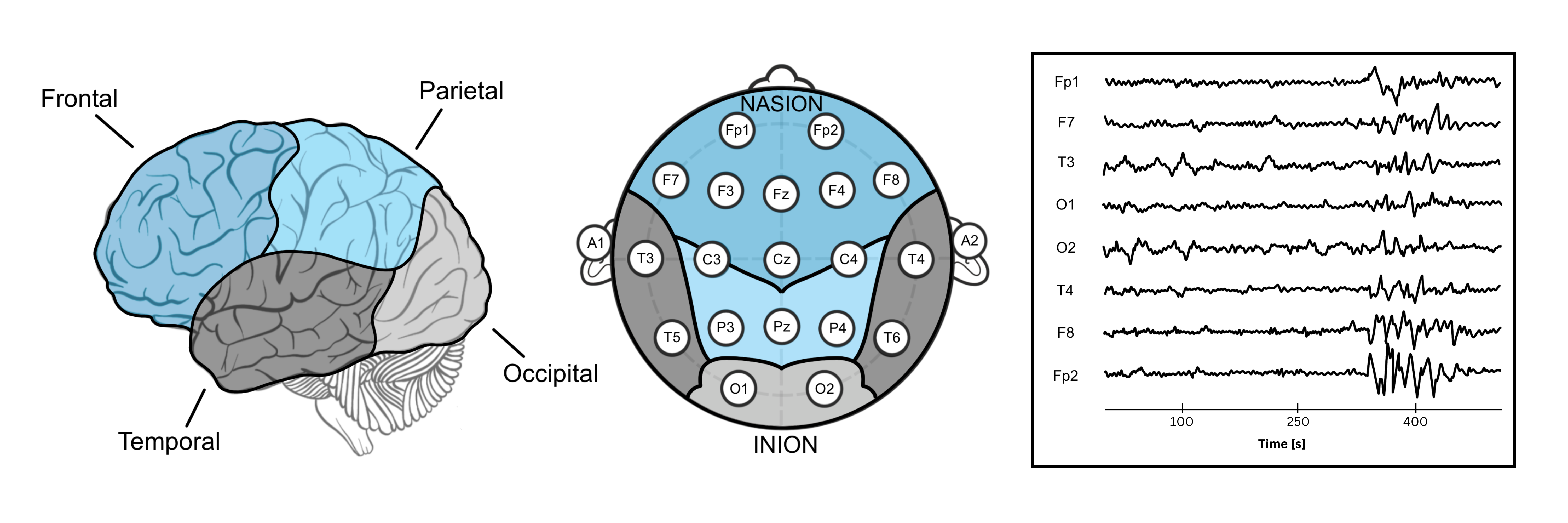}
\caption{Brain regions and placement of scalp electrodes according to the 10--20 system, along with (simulated) examples of EEG signals.}
\label{fig:eeg-illustration}
\end{figure}
Figure~\ref{fig:eeg-illustration} displays the locations of scalp electrodes in the human brain following the international 10--20 system for EEG \citep{eeg1020}. The labeling letters indicate the brain region covered by the electrodes. From front to the posterior part: Fp (pre-frontal or frontal pole), F (frontal), C (central line of the brain), T (temporal), P (parietal), and O (occipital). Sites marked with a ``z'' (zero) refer to an electrode placed on the midline sagittal plane of the skull (e.g., Fz, Cz, Pz). Electrodes labeled with an A are the earlobes, typically serving as an offline reference for signal analysis. These electrodes sometimes are placed on the mastoids, the bone behind the ear. Odd numbers represent electrodes on the left lateral, and even numbers represent electrodes on the right lateral; these numbers increase, as the electrodes move further from the midline. In addition, Figure~\ref{fig:eeg-illustration} shows some simulated EEG signals according to a 10--20 system, illustrating the typical behavior of EEGs.

\subsection{Multivariate extreme value theory}
Let $\bm{X}_d := \left(X_{d}^{1}, \ldots, X_{d}^{T}\right)^{\intercal}$, $d=1,\ldots,D$, be the amplitudes from the $d$-th EEG channel. We assume that $\bm{X}_d$ is a univariate stationary time series. We further assume the existence of real sequences $a_{d}^{T} > 0$ and $b_{d}^{T} \in \Rset$, for each $d$, such that
\begin{equation}
\Pr\left( \frac{\max{\bm{X}_{1}} - b_{1}^{T}}{a_{1}^{T}}\leq x_1, \ldots, \frac{\max{\bm{X}_{D}} - b_{D}^{T}}{a_{D}^{T}}\leq x_D\right) \stackrel{T\to \infty}{\longrightarrow} G(x_1, \ldots, x_D),
\label{eq:mevt}
\end{equation}
where $\max{\bm{X}_{d}} = \max{(X_{d}^{1}, \ldots, X_{d}^{T})}$ is the componentwise maximum and $G$ is, in each margin, a non-degenerate distribution function.

According to the multivariate version of the extremal types theorem \citep{deHaan_book}, $G$ is a multivariate extreme-value distribution with generalized extreme value (GEV) margins. Specifically, each marginal distribution of the amplitude of the EEG signal at each channel $d$ is given by
$$G_{d}(x;\, \mu_d, \sigma_d, \xi_d) = \exp\left[-\left(1 + \xi_{d}\frac{x - \mu_d}{\sigma_d} \right)^{-1/\xi_d}\right],\, \text{for}\,\, 1 + \xi_{d}\frac{x - \mu_d}{\sigma_d} > 0, $$
where $\mu_d \in \Rset$ is the location parameter, $\sigma_d > 0$ is the scale parameter, and $\xi_d \in \Rset$ is the extreme value index, a parameter characterizing the extreme behavior of component $d$. The distribution $G_d$ belongs to either the Gumbel ($\xi_{d}=0$), the Fréchet ($\xi_{d}>0$) or the (reversed) Weibull ($\xi_{d}<0$) family. 

Now, let $F_{d}(x) = \Pr(X^{t}_{d} \leq x)$ be the marginal distribution function of the stationary time series $\bm{X}_d$, $d = 1, \ldots, D$, here assumed to be continuous. The convergence in Eq.~\ref{eq:mevt} holds if, and only if, for any $t=1, \ldots, T$, the vector
\begin{equation}
\bm{Y}^t = \left( \left[1 - F_{1}(X_{1}^{t})\right]^{-1}, \ldots,  \left[1 - F_{D}(X_{D}^{t})\right]^{-1}\right)^{\intercal},
\label{eq:yvector}
\end{equation}
satisfies the following multivariate regularly varying assumption
\begin{equation}
\dfrac{\bm{Y}^t}{\normx{\bm{Y}^t}_{2}} \, \bigg| \, \normx{\bm{Y}^t}_{2} > \tau \,\, \rightharpoonup \,\, S,\quad \tau \to \infty,
\label{eq:mrv}
\end{equation}
for a giving high threshold \(\tau\) and a random vector \(S\) with probability measure \(\mathcal{S}\) on the restricted unit sphere 
\(\mathbb{S}_{+}^{D-1} = \{\bm{s}\in\mathbb{R}_{+}^{D}:\|\bm{s}\|_{2}=1\}\). Here, \(\|\cdot\|_{2}\) denotes the Euclidean norm (although other norms, such as the \(L_1\) or sup‐norm, could be used equally well without changing the core theory) and “\(\xrightarrow{}\)” denotes weak convergence on \(\mathbb{S}_{+}^{D-1}\). By definition, $\bm{Y}^t$ has unit Pareto margins; hence, if it satisfies Eq.~\ref{eq:mrv}, then it is in the maximum domain of attraction of a max-stable distribution with Fréchet margins and is uniquely specified by the angular measure $\mathcal{S}$. The angular measure \(\mathcal{S}\) thus characterizes the extremal dependence structure (or copula) of both:
\begin{itemize}
\item \textbf{Block maxima}, obtained by dividing the observation period into non-overlapping intervals of equal length and retaining the maximum in each; and
\item \textbf{Threshold exceedances}, obtained by selecting those observations that exceed a sufficiently high threshold
\end{itemize}
of the multivariate standardized vector \(\bm{Y}^t\), and therefore also describes the extremal behavior of \(\bm{X}^t\) on the original scale.

\subsection{Spherical \texorpdfstring{$k$}--means clustering algorithm}
Our goal is to cluster EEG channels whose large amplitudes in their signals are synchronized---either contemporaneous or with some time delay. To do so, we need to define the concept of asymptotic dependence (AD) and asymptotic independence (AI) between pairs of EEG signals. In the bivariate setting, the primary measure of asymptotic (in)dependence between two variables $Y_d$ and $Y_{d^{\prime}}$ defined through Eq.~\ref{eq:yvector} is the tail correlation coefficient computed as
\begin{equation}
\chi_{dd^{\prime}} := \lim\limits_{\tau \to \infty}\Pr\left(Y_{d^{\prime}} > \tau \mid Y_d > \tau\right) = \frac{1}{\mu}\E\left(S_d \wedge S_{d^{\prime}}\right)\in [0,1],\, d,d^{\prime}=1,\ldots,D,
\label{eq:chichi}
\end{equation}
where $\mu=\E\left(S_d\right)$, $d = 1, \ldots, D$, and ``\(\wedge\)'' denotes the minimum operator, i.e., \(S_d \wedge S_{d^{\prime}} \;=\; \min\bigl(S_d, S_{d^{\prime}}\bigr)\). The case $\chi_{dd^{\prime}} = 0$ corresponds to AI while $\chi_{dd^{\prime}} > 0$ corresponds to AD. The link to the angular measure reveals that it determines the occurrence of joint extremes. Suppose $\mathcal{S}$ has mass on a particular face of the sphere. In that case, this implies a positive probability that the corresponding variables experience extremes in synchrony (i.e., they are AD). In contrast, if $\mathcal{S}$ has no mass on a specific face, this implies AI for the corresponding variables. Our goal is to build a clustering algorithm such that channels are mutually AD within clusters (i.e., we are likely to observe high amplitudes of all channels in synchrony) but AI across them.

To achieve our goal, we use a clustering procedure able to detect regions, in the faces of $\Sset_{+}^{D-1}$, where the mass of the angular measure is concentrated. That is, we need to group extreme observations with similar angles determined by the support of the measure $\mathcal{S}$. There are two well-studied options in the literature: the spherical $k$--means (s--$k$--m) clustering algorithm by \cite{janssen21} and the spherical $k$--principal--component (s--$k$--pc) clustering algorithm by \cite{fomichov22}. In the former approach, similar observations are grouped into $k$ distinct clusters by minimizing a cosine dissimilarity \(\varphi(\bm x, \bm y) = 1 - \cos{(\bm x, \bm y)} = 1 - {\bm x}^{\intercal}\bm y\), \(\bm x,\bm y\in\Sset_{+}^{D-1}\), using an adaptation of the s--$k$--m algorithm by \cite{dhillon2001}. Hence, the identified cluster centers can be thought of as angular prototypes for extremes, i.e., they reveal which angles of the extremes are the most likely. The latter is a two-step approach where a principal component analysis (PCA) suited for extremes \citep{drees2021} is combined with clustering. The first principal eigenvector is used to reduce the data dimensionality; then, a $k$-means algorithm with a quadratic dissimilarity \(\varphi_{2}(\bm x, \bm y) = 1 - ({\bm x}^{\intercal}{\bm y})^2\), \(\bm x,\bm y\in\Sset_{+}^{D-1}\), is applied to identify the $k$ clusters.

Although, in practice, both approaches lead to similar results, the s–\(k\)–m method is computationally faster for complex high-dimensional data with a potentially larger number of clusters \citep{chengthesis}. Specifically, the s–\(k\)–pc algorithm has running‐time complexity on the order of \(\mathcal{O}\bigl(k\,(T\,d + d^{2}\log d)\bigr)\), whereas the s–\(k\)–m algorithm requires only \(\mathcal{O}(T + k)\) operations. This computational advantage is why, in this work, we use the s–\(k\)–m algorithm.

In summary, we therefore aim to find $k$ centroids $\bm{a}_c \in \Rset_{+}^{D}$, $c\in \{1, \ldots, k\}$, such that the expected cosine dissimilarity of extreme observations to their closest cluster centroid is minimal. Hence, let $\{\bm{Y}^t\}_{t=1}^{T}$ be the multivariate time series of standardized signals amplitudes given by Eq.~\ref{eq:yvector}. Assume we have $\Tau$ extreme observations above for which the L2 norm exceeds the threshold $\tau$ and let $\bm{\Theta} = \{\bm{\theta}^1, \ldots, \bm{\theta}^{\Tau}\}$ be the pseudo-angles, or angular components, of such observations, i.e., the normalized vectors 
\(\bm{\theta}^t = \bm{Y}^t / \|\bm{Y}^t\|_2\) lying on the positive unit sphere. These pseudo-angles have an approximate distribution $\mathcal{S}$ given by Eq.~\ref{eq:mrv}. By construction, each $\bm{\theta}^{t^{\prime}}$ satisfies $\|\bm{\theta}^{t^{\prime}}\|_2 = 1$ and hence lies in the positive unit sphere $\Sset_{+}^{D-1}$, for all $t^{\prime}\in\{1, \ldots, \Tau\}$. Let $A = \{\bm{a}_1, \ldots, \bm{a}_k\}\subset\Sset_{+}^{D-1}$ be the set of cluster centroids. Since the extreme amplitudes are constrained to live on the unit sphere, the cosine dissimilarity can be used in its simplified version $\phi(\bm{\theta}^{t^{\prime}}, \bm{a}_c) := 1 - \cos{({\bm{\theta}^{t^{\prime}}}, \bm{a}_c)} = 1 - {\bm{\theta}^{t^{\prime}}}^{\intercal}\bm{a}_c$, which is the spherical distance of the $t$-th extreme pseudo-angle of $\bm{\Theta}$ to the centroid $\bm{a}_c \in A$. Hence, the minimized distance from arbitrary extreme observations, with angles $\bm{\theta}^{t^{\prime}}$, to their nearest cluster centroid on $A$ is $\varphi(\bm{\Theta}, A):={\Tau}^{-1}{\sum\nolimits_{{t^{\prime}}=1}^{\Tau}{\min\nolimits_{\bm{a}_c \in A}\phi(\bm{\theta}^{t^{\prime}}, \bm{a}_c)}}$. Given $\bm{\Theta}$, we denote the optimal set of size $k$ minimizing $\varphi(\bm{\Theta}, \cdot)$ as $A_{k}^{\Tau}$. The entire spherical clustering algorithm, embedded in the Club Exco method, is fully detailed in Section~\ref{chp:proc}.

\begin{figure}[!hbt]
\centering
\includegraphics[width=\textwidth]{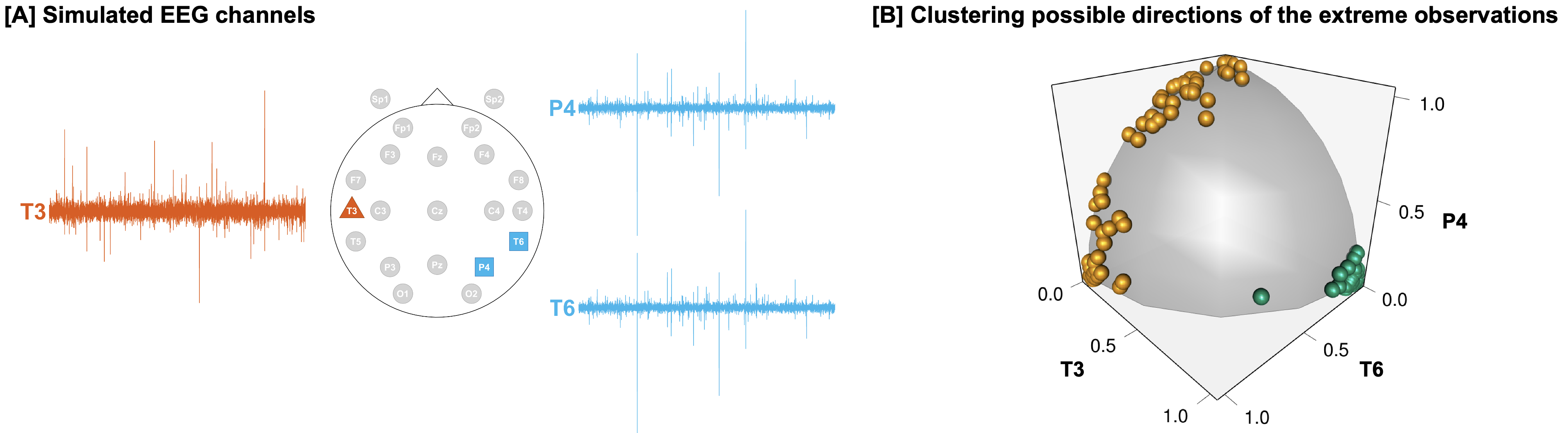}
\caption{In panel A, simulated EEG signals from two AD channels (P4 and T6) and from one channel (T3) that is AI with the others.  In panel B, extreme observations from the triplet (T3, P4, T6) are projected onto the unit positive octant suggesting AD between P4 and T6 but not with T3.}
\label{fig:clustering}
\end{figure}

To illustrate how Club Exco works, Figure~\ref{fig:clustering}, panel A, shows three simulated EEG signals. Channels P4 and T6 are simulated under the moving average process of order 3 (MA(3)) $X^{t} = Z^{t} + 0.5Z^{t-1} - 0.6Z^{t-2} + 1.5Z^{t-3}$, where $\{Z^{t}\}_{t=1}^{T}$ is an independent and identically distributed sequence of symmetric stable random variables with index $\alpha = 1.75$ \citep{nolanbook}. For illustration, set P4 and T6 as the time series given by $X^t$ and its lag-1 version, $X^{t-1}$, respectively. Hence, P4 and T6 are AD with $\chi \approx 0.28$, calculated empirically from Eq.~\ref{eq:chichi} by taking the threshold $\tau$ as the empirical $0.998$--marginal quantile. 
Channel T3, on the other hand, is simulated, independently from P4 and T6, using the MA(4) process $X_{\star}^{t} = Z_{\star}^{t} + 0.7Z_{\star}^{t-1} - 0.2Z_{\star}^{t-2} + 1.5Z_{\star}^{t-3} - 0.5Z_{\star}^{t-4}$, where $Z_{\star}$ is obtained similarly as $Z$ with the same index $\alpha_{\star} = \alpha = 1.75$.

Therefore, T3 is trivially AI with both P4 and T6. Figure~\ref{fig:clustering}, panel B, displays the 1\% most extreme amplitudes of the triplet (T3, P4, T6), as determined by their L2 norm, projected onto the unit positive octant. We apply the s--$k$--m algorithm with $k=2$ clusters to group these extreme observations. The two estimated clusters (gold and green points in panel B) suggest that channels P4 and T6 are indeed AD because the corresponding points (gold) on the sphere octant point toward joint extremes in both P4 and T6 but separately from T3. By contrast, channel T3 appears indeed to be AI with P4 and T6 because the correspondent points (green) point at extremes of T3 alone. Basically, the angles of the vectors pointing to the center of the gold and green points define the clusters’ centroids.

\subsection{Club Exco procedure\label{chp:proc}}
The application of Club Exco comprises two steps delineated in Algorithms~\ref{alg:clubexco} and \ref{alg:skm}. Algorithm~\ref{alg:clubexco} functions as a standard preprocessing step in multivariate extreme value analysis, with the objective of identifying the pseudo-angles of extreme observations from the data. Algorithm~\ref{alg:skm} employs the s--$k$--m algorithm on the pseudo-angles to obtain the cluster centroids. Furthermore, one can choose to work with either the amplitudes of the original time series or the bandpass-filtered time series to investigate frequency-specific tail dependence patterns between channels.

\RestyleAlgo{boxruled}
\NoCaptionOfAlgo
\begin{algorithm}[ht]{
{\footnotesize{
\caption{\textbf{Club Exco Algorithm 1: obtaining extreme pseudo-angles.}\label{alg:clubexco}}

\KwData{A $T$-by-$D$ matrix $\bm{X} = [\bm{X}_1,\ldots, \bm{X}_D]$ where columns contain the signal amplitudes from the EEG channels or bandpass-filtered time series.}

\vspace*{.15cm}

\KwResult{Extreme pseudo-angles $\bm{\Theta} = \{\bm{\theta}^{1},\ldots, \bm{\theta}^{\mathcal{T}}\}$.}
  
\BlankLine
 
\Begin{

\textit{\textbf{1.}} Transform data on the original scale $\bm{X}$ to the unit Pareto scale $\bm{Y}$:
$$\bm{Y}^t \gets \left( \left[1 - F_{1}(X_{1}^{t})\right]^{-1}, \ldots,  \left[1 - F_{D}(X_{D}^{t})\right]^{-1}\right),\; t = 1, \ldots, T.$$

\textit{\textbf{2.}} For all $t\in\{1,\ldots, T\}$, keep the transformed observations with largest Euclidean norm: 
$$\left\{ {\bm{Y}^{*}}^{t^{\prime}}\right\}_{t'=1}^{\mathcal{T}} \gets \left\{\bm{Y}^{t}\, \big|\, \normx{\bm{Y}^t}_{2}>\tau\right\}_{t=1}^{T}.$$
  
\textit{\textbf{3.}} Project the transformed observations onto the unit sphere:
$$\bm{\theta}^{t^\prime} \gets  \frac{{\bm{Y}^{*}}^{t^\prime}}{\normx{{\bm{Y}^{*}}^{t^\prime}}_{2}},\; t^{\prime} = 1, \ldots, \mathcal{T}.$$
}
}}}
\end{algorithm}

\RestyleAlgo{boxruled}
\NoCaptionOfAlgo
\begin{algorithm}[ht]{
{\footnotesize{
\caption{\textbf{Club Exco Algorithm 2: applying spherical $\bm{k}$--means.}\label{alg:skm}}

\KwData{Extreme pseudo-angles $\bm{\Theta} = \{\bm{\theta}^{1},\ldots, \bm{\theta}^{\mathcal{T}}\}$ and initial cluster centroids $A = \{\bm{a}_1,\ldots, \bm{a}_k$\}.}

\vspace*{.15cm}

\KwResult{Updated cluster centroids $A^{*} = \{\bm{a}^{*}_1,\ldots, \bm{a}^{*}_k$\} and minimized distance $\varphi(\bm{\Theta}, A^{*})$.}

\BlankLine
 
\Repeat{convergence}{
\textit{\textbf{1.}} Assign $\bm{\theta}^{t^{\prime}}$ to the cluster with closest centroid $\bm{a}_c$, $c=1,\ldots, k$, by computing 
$$\arg\min\nolimits_{\bm{a}_c \in A}\phi(\bm{\theta}^{t^{\prime}}, \bm{a}_c),\, {t^{\prime}} = 1, \ldots, \mathcal{T}.$$

\textit{\textbf{3.}} Update cluster centroids, $A$.

\vspace*{.275cm}

\textit{\textbf{2.}} Update the overall distance measure:
$$\varphi(\bm{\Theta}, A) \gets {\Tau}^{-1}{\sum\nolimits_{{t^{\prime}}=1}^{\Tau}{\min\nolimits_{\bm{a}_c \in A}\phi(\bm{\theta}^{t^{\prime}}, \bm{a}_c)}}.$$

}
}}}
\end{algorithm}

\newpage
\section{Clustering multi-channel EEG seizure data\label{sec:application}}
This section applies the Club Exco method to neonatal EEG recordings. First, we detail the seizure versus non-seizure cohort and our preprocessing pipeline. Next, we employ a sliding-window spherical $k$-means clustering on extreme amplitude events to identify evolving brain extreme communities. Finally, we benchmark these findings against coherence-based clusters obtained using the HCC method.

\subsection{Dataset, Preprocessing and Scientific Objectives}

In this section, we analyze neonatal EEG recordings, collated by \cite{stevenson2019dataset}, from infants admitted at the neonatal intensive care unit (NICU) of Helsinki University Hospital under the suspicion of experiencing seizures. The dataset includes about an hour of EEG recordings (at a sampling rate of 256 Hz), observed from the standard 10--20 system 19-channel EEG electrodes with a reference electrode at the midline, for each of 79 neonatal patients. Three expert clinicians independently annotated each subject's data, labeling neonates as having experienced a seizure or as ``seizure-free,'' meaning no seizures were observed during any monitored period in the NICU. Consensus among the experts resulted in the classification of 39 patients as having experienced ictal events and 22 patients as having no observed seizures. For the remaining 18 patients, no consensus was reached among the three experts. Throughout the remainder of this paper, we refer to the first two groups as ``seizure patients'' and ``non-seizure patients,'' respectively.

To remove unwanted artifacts while ensuring the extreme behavior of the signals due to seizure remains unchanged, each recording underwent a data cleaning protocol. Specifically, we selected a subset channels that have minimum level of quality in all subjects, resulting in the exclusion of channels Fz and Cz in the analysis. Then, we applied a preprocessing technique, namely, the Early Stage Preprocessing (PREP) pipeline of \cite{bigdely2015prep}, to enhance the quality of the signals. Afterwards, we chose not to include data which exhibit suspiciously small values in the analysis. Considering only patients with a complete set of available clinical information, the pre-processed data we finally analyze involves 13 non-seizure patients and 17 seizure patients whose annotations were confirmed by all three specialists.

In our data analysis, the main goals are: (1.) to identify channels sharing the same extreme community; (2.) to study how community memberships evolve during seizures; and (3.) to differentiate the brain connectivity patterns between patients suffering from seizures and those that are seizure-free. In addition, we are interested in verifying whether extreme communities (based on the tails of the probability distribution) differ from clustering procedures based on the entire probability distribution, such as the HCC by \cite{euanHCC}, in this context and whether this brings additional clinical insights.

\subsection{Application of the Club Exco Method}

\begin{figure}[!bht]
\centering
\includegraphics[width=\textwidth]{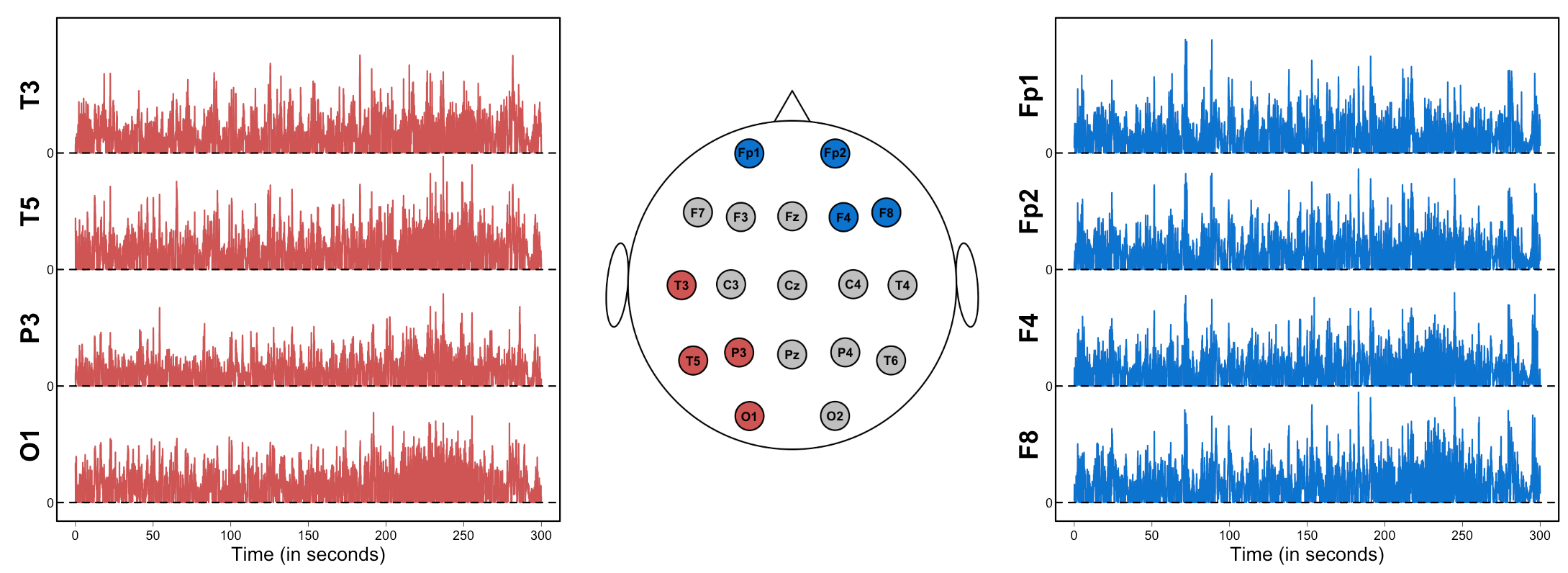}
\caption{Selected EEG amplitudes during seizure of a neonatal patient. Channels highlighted in blue exhibit synchronized extreme behavior patterns (right) while channels highlighted in red produce seemingly asymptotically independent signal amplitudes (left).}
\label{fig:signals}
\end{figure}

The Club Exco method is tailored to address our first interest. It enables for identifying groups of EEG channels for which high amplitudes in the signals are much more likely to occur together (with potential lead-lag between channels), i.e., estimating clusters of asymptotically dependent channels. We refer to these clusters as brain extreme communities. Figure~\ref{fig:signals} presents 5-minute recordings from eight selected EEG channels of a neonatal patient that suffers from seizure. Channels in the frontal region, namely channels Fp1, Fp2, F4 and F8, are examples of nodes in a brain network that belong in the same brain extreme community. Large amplitude signals from these channels seem to occur together, i.e., they exhibit asymptotic dependence, which is typically expected during a seizure episode. By contrast, extremes of the signal amplitudes from channels T3, T5, P3 and O1 appear to be less synchronous in the upper tail. This may be an indication that these nodes belong to different brain extreme communities or are simply asymptotically independent. 

For each neonatal patient, to investigate how community membership of the channels evolve across different seizure occurrences, we employ a sliding window approach where we apply our Club Exco method separately over $W = 124$ time windows (of 10-second duration with 2-second overlap with adjacent windows) to estimate brain extreme communities. More precisely, we consider estimating $k$ clusters assuming the threshold $\tau$ (see Eq. \ref{eq:mrv} and Eq. \ref{eq:chichi}) as the $\tau$-quantile of the signal amplitudes for each time window from a specific patient (Algorithm 1, step 2). Then, we calculate the proportion of a pair of channels belonging in the same brain community across all time windows for all possible channel pairs. Thus, we obtain what we call the ``extreme connectivity persistence (ECP)'' matrix. We should note that persistence in the (non-extreme) connectivity patterns has been used as predictive biomarkers where persistent and high strength connectivity is associated with positive treatment response in epileptic spams \citep{StrengthStabilityEEG-Shrey-2018}.

ECP as a metric allows us to identify groups of channels that are asymptotically dependent with each other (which reflects the impact of seizures), and channels that are asymptotically independent from such groups.
Larger values for $k$ only lead to splitting an asymptotically dependent cluster into smaller groups of channels, resulting in smaller proportions for the ECP matrices (see Supplementary Material). Thus, we find estimating $k=2$ clusters to be appropriate for our objective, which provides better insights and a more practical perspective in the extreme brain connectivity during seizures. Furthermore, we choose $\tau = 0.9$ because it ensures that the amplitudes are large enough, i.e., they appropriately represent the tail behavior of the signals, while maintaining a sufficient number of observations available for estimation.

\begin{figure}[!th]
\centering
\includegraphics[width=0.425\textwidth]{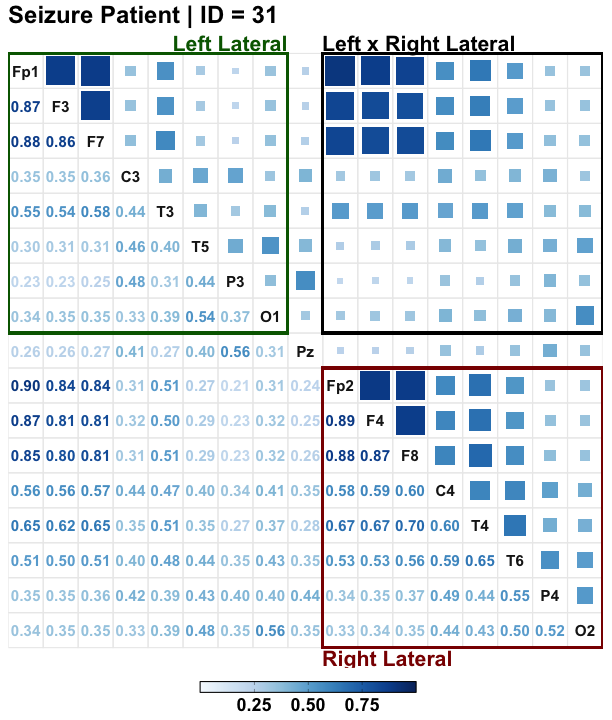}
\includegraphics[width=0.425\textwidth]{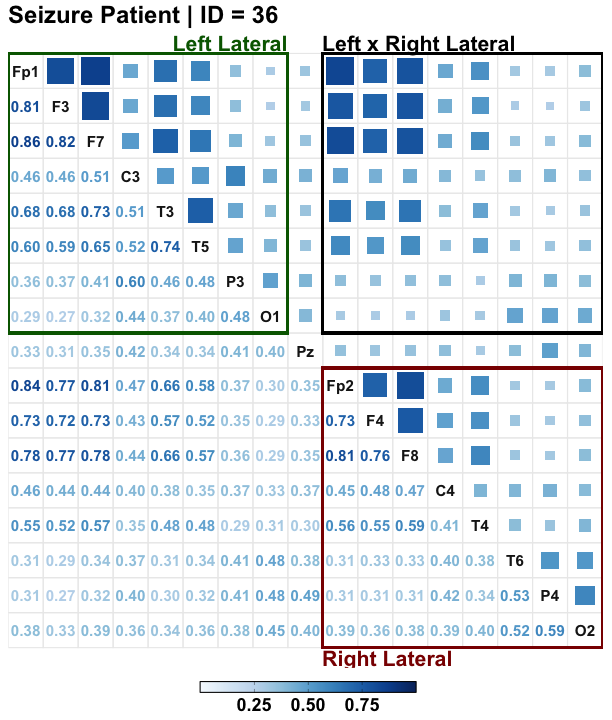}

\includegraphics[width=0.425\textwidth]{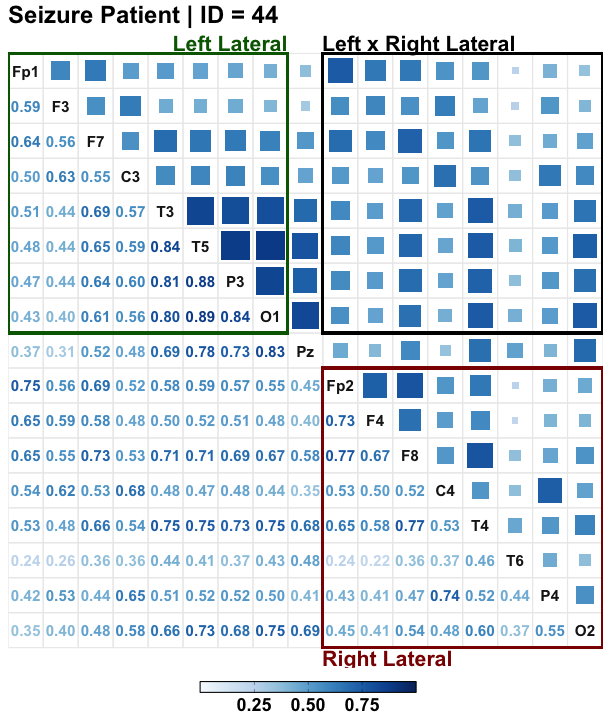}
\includegraphics[width=0.425\textwidth]{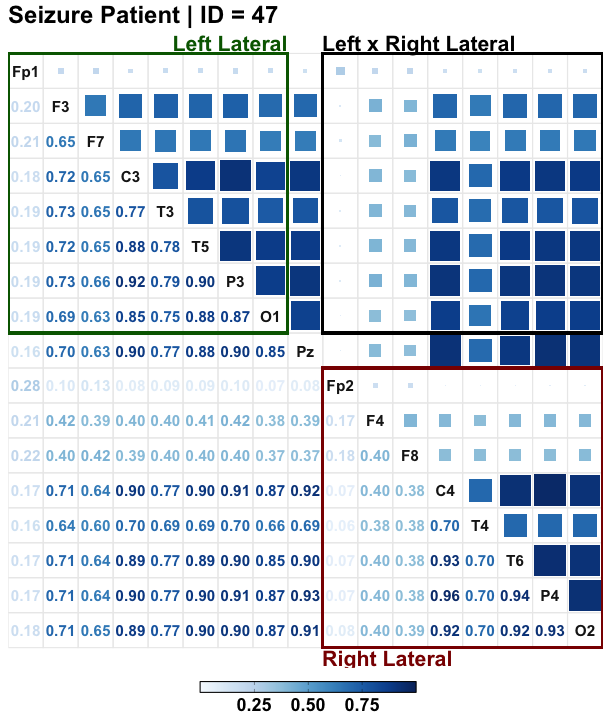}
\caption{Estimated extreme connectivity persistence matrices for four selected neonatal patients (with ID $= 31, 36, 44$, and $47$) suffering from seizures. In each plot matrix, the numbers (lower triangle) are the actual proportions of cluster persistence, the shaded squares (upper triangle) visualize the proportions where larger and darker shade represent larger magnitude, and the boxes highlight the channels in the left lateral (upper left) and right lateral (lower right) regions, and the cross-lateral interactions (upper right).}
\label{fig:permat_subj}
\end{figure}

Figure~\ref{fig:permat_subj} illustrates the ECP matrices for three neonatal patients annotated to have experienced actual seizures. For example, the first off-diagonal entry from the patient with ID $= 31$ indicates that the adjacent channels Fp1 and F3 belong in the same brain extreme community for about 87\% of all time windows considered. That is, large amplitudes from these two channels persistently co-occur across the observed seizure occurrences. By contrast, the non-adjacent channels Fp2 and P3 are seldom clustered together, which are detected to belong in the same community in only about 21\% of all time windows and thus, co-occurrence between large amplitudes of these channels are deemed to be less persistent. Hence, constructing these ECP matrices offers an exploratory tool that enables for understanding how brain connectivity, through clustering via extremal behavior, evolves as seizure manifests.

\begin{figure}[!ht]
\centering
\includegraphics[width=0.9\textwidth]{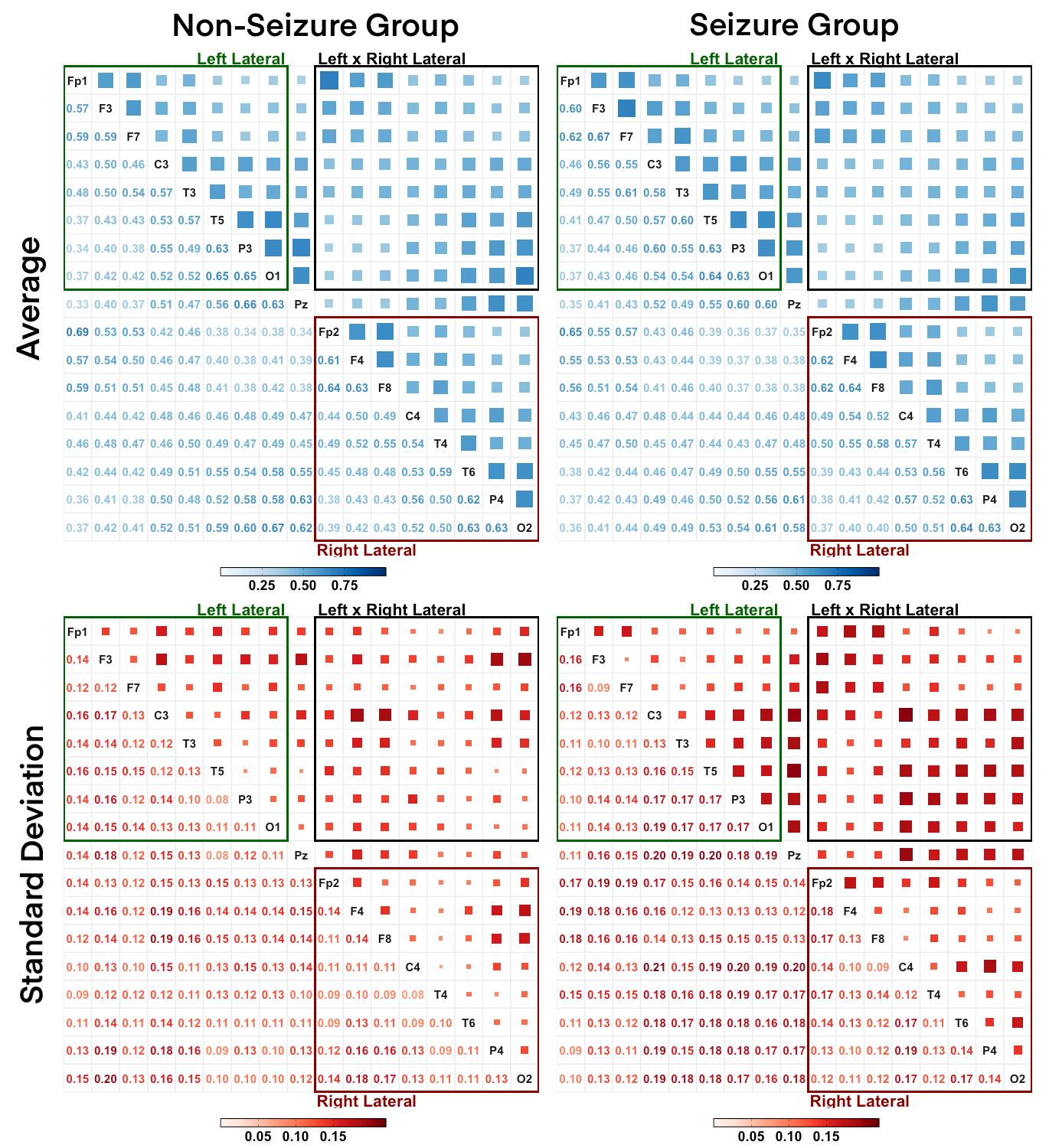}
\caption{Average (top) and standard deviation (bottom) of extreme connectivity persistence matrices among non-seizure patients (left) and patients suffering from seizures (right) assuming $k=2$ clusters at threshold $\tau = 0.9$. In each plot matrix, the numbers (lower triangle) are the actual values, the shaded squares (upper triangle) visualize the values where larger and darker shade represent larger magnitude, and the boxes highlight the channels in the left lateral (upper left) and right lateral (lower right) regions, and the cross-lateral interactions (upper right).}
\label{fig:permat_summ}
\end{figure}

To facilitate interpretation, we summarize all ECP matrices for seizure patients and non-seizure patients in Figure~\ref{fig:permat_summ} (through averages and standard deviations across patients for each group), and highlight few individual cases to illustrate the differences in brain connectivity between the two groups in Figure~\ref{fig:permat_subj}. All individual results for each patient, including a sensitivity analysis assuming different number of clusters (i.e., $k = 3, 4, 5,$ and $6$), are reported in the Supplementary Material due to space limitations. In Figure~\ref{fig:permat_summ}, we observe highly similar average extreme connectivity patterns between the two groups of patients. These persistent clusters denote a robustness of the ECP matrices to detect communities.

Precisely, adjacent channels tend to cluster together, e.g., the frontal channels exhibit persistent asymptotic dependence as well as the temporal-parietal-occipital (TPO) channels. We expect this to be the case as EEG signals from channels that are close to each other may have originated from the same neuronal population and hence, share similar extremal behavior in their amplitudes by default. However, one notable difference between the groups is that the persistence of clustering spreads slightly more to non-adjacent channels for patients in the seizure group. This is highlighted by the relatively higher average proportions between the frontal and TPO channels, especially in the left lateral region. For instance, channel F7 belongs to the same extreme brain community as channels T5 and P3, on average, more often among the seizure patients (with persistence proportion of 50\% and 46\%, respectively) than among the non-seizure patients (with persistence proportion of 43\% and 38\%, respectively).

Another striking difference between seizure and non-seizure patients is the variability in extreme connectivity persistence within groups. Subject-specific connectivity patterns have also been observed in correlation-based connectivity in patients with infantile spasms \citep{EffectInterictalEpileptiform-Hu-2020}, and binary correlation connectivity patterns \citep{ConnectivityHighFreq-PintoOrellana-2024}. But it has not been observed in extreme value dependence, as it is observed that spikes (most common physiological events related to extremes) do not affect functional connectivity \citep{DynamicalMechanismsInterictal-Courtiol-2020}.
Furthermore, some metrics derived from EEG networks are more unstable during ictal intervals than inter-ictal periods \citep{NeuralFragilityEEG-Li-2021}.

Our results shows that for non-seizure patients, cluster membership of adjacent channels exhibit less variability while channels that are far apart cluster in a more variable manner (see Figure~\ref{fig:permat_summ}, bottom panels). In both left and right lateral regions, frontal channels (i.e., channels Fp1, F3 and F7 in the left, and channels Fp2, F4 and F8 in the right) belong to the same brain community more persistently and more consistently, which is also true for TPO channels (i.e., channels T3, T5, P3 and O1, and channels T4, T6, P4 and O2). Moreover, cluster memberships between frontal channels and TPO channels are less persistent and less consistent, whether via intra-lateral (e.g., left with left lateral) or cross-lateral (e.g., left with right lateral) interactions. This coincides with the natural characteristics of the EEG signals where recordings associated with the same neuronal population tend to exhibit similar extreme amplitude behaviors. Moreover, we observe the opposite for the seizure patient group. The variability of extreme connectivity persistence is larger for adjacent channels, e.g., cluster membership between TPO channels in the left lateral region varies greatly, and persistence of clustering between cross-lateral adjacent channels exhibit larger variability. Precisely, frontal channels in the left and right lateral regions cluster more persistently yet less consistently, which is also the case for the TPO channels in the two lateral regions.

We speculate that these differences may be attributed to how seizure manifests in the brain. In Figure~\ref{fig:permat_subj}, the estimated ECP matrices from the selected neonatal patients highlight how brain connectivity is affected during seizure, which we describe as ``localization'' and ``propagation''. For seizure patients with ID $=31$ and ID $=36$, the left frontal channels exhibit highly persistent clustering among each other while cluster membership to the same brain extreme community between the remaining channels are less persistent. We associate this to the localization of seizure where a cluster of channels exhibit persistent co-occurrence of extreme amplitudes while others channels, regardless if they are adjacent or non-adjacent channels, exhibit asymptotic independence. We should note that patient ID$=31$ and ID$=36$ have been diagnosed with seizures localized primarily in the right and both hemispheres, respectively \cite{stevenson2019dataset}. These results suggest us that the extreme seizure dynamics may be associated with synchronization processes outside the seizure onset zone as showed by Kini el al. \citep{VirtualResectionPredicts-Kini-2019}.

By contrast, we observe seizure propagation for patient with ID $=47$. Clustering among the left and right TPO channels are highly persistent as seizure propagates to the left frontal (i.e., channels F3, F7 and C3) and right (i.e., channel C4) region of the brain. For patient with ID $=44$, the estimated ECP matrix suggests that seizures have both localized and propagated effects. Specifically, the left TPO channels are persistently clustered in the same brain extreme community, while the right TPO channels are seldom clustered together. Moreover, the right frontal channels also exhibit persistent clustering, sharing the same community as the left TPO channels, which we associate to the diagonal cross-lateral seizure propagation. Both patients have been diagnosed with seizures in both hemispheres \cite{stevenson2019dataset}, and their propagation patterns seemed to have persistent contra-lateral communities.

Even though every patient has a unique ECP pattern, the average values in the seizure and non-seizure groups (involving 17 and 13 patients, respectively) share a similar structure. This default pattern suggests that a physiological process, likely sleep, may be involved in the dynamics of extremes, characterized in our results by predominant contralateral synchronization between the frontal and occipital regions. This finding could be related to one of the newborn cortical states identified by \cite{NeonatalCorticalActivity-Khazaei-2023}, where synchronization patterns in the frontal region frequently exhibit the highest connectivity strength across 47 hypoxic-ischemic encephalopathy newborns. Furthermore, our findings in the occipital region may be related to one of the phenomena observed during quiet sleep, which has been linked to periods of hypoactivity and bursts of high amplitudes in this region, described by \cite{RvwSleepEEG-Dereymaeker-2017}, refer to as ``occipital delta brushes''. Club Exco provides a robust approach to quantify synchronization patterns of extreme values in brain signals. However, due to limitations inherent to our dataset, we are unable to provide a more rigorous interpretation without a robust control group for comparison.

\subsection{Complementary Analysis via the HCC Method}

In neuroscience, a standard practice is to decompose brain activity into five canonical bands: the delta (0–4 Hz), theta (4–8 Hz), alpha (8–12 Hz), beta (12–30 Hz), and gamma (30–50 Hz) bands \citep{ombao2005slex, nunez2016electroencephalography, fiecas2016modeling, redondo2022functional, guerrero23, redondo2023measuring} or even, higher intervals as ripple (80-200 Hz) and fast-ripple (200-500) bands \citep{ConnectivityHighFreq-PintoOrellana-2024}. This is because each frequency band reflects distinct cognitive functions or physiological processes. For instance, the alpha band is linked to relaxed wakefulness, the beta band to external attention, and the gamma band to concentration \citep{abhang16bookchp2}. For a detailed overview of frequency-domain EEG analysis, see \cite{Ombao2024SpectralDependence}. Given each band's differing neurological roles, clustering based on the characteristics of these frequency oscillations may help understand the complex mechanism associated to seizures.

Unlike our Club Exco method which utilizes EVT to form clusters based on tail dependencies, the HCC method of \cite{euanHCC} uses a hierarchical clustering algorithm that derives the dissimilarities between clusters of channels from the coherence measure (an analog of the cross-correlation between two time series at different frequency oscillations). Although coherence is a measure of dependence that accounts for the entire data distribution, i.e., it does not focus on specific data ranges such as extremes from the upper tail, an advantage of the HCC method is that it provides an understanding of the clustering dynamics in the frequency domain. Hence, we complement our analysis by employing a similar approach to summarize the clustering results with the HCC method.

More precisely, we estimate cluster membership via the HCC method for each time window, assuming $k = 2$ clusters, from the EEG recordings of neonatal seizure and non-seizure patients. Then, we calculate the proportion of windows where a pair of channels belong to the same brain community, which we here call the ``spectral connectivity persistence (SCP)'' matrix. In addition, we focus our attention to two sets of frequencies, namely, the low frequencies consisting of the delta, theta and alpha bands, and the high frequencies including the beta and gamma bands. By looking into these set of frequency bands, our aim is to understand how seizure occurrences affect brain connectivity in the frequency domain and establish its link to cognitive neurological interpretations. Lastly, we aggregate the SCP matrices for each set of frequency bands across all neonates from the seizure patient group and from the non-seizure patient group.

\begin{figure}[!th]
\centering
\includegraphics[width=0.9\textwidth]{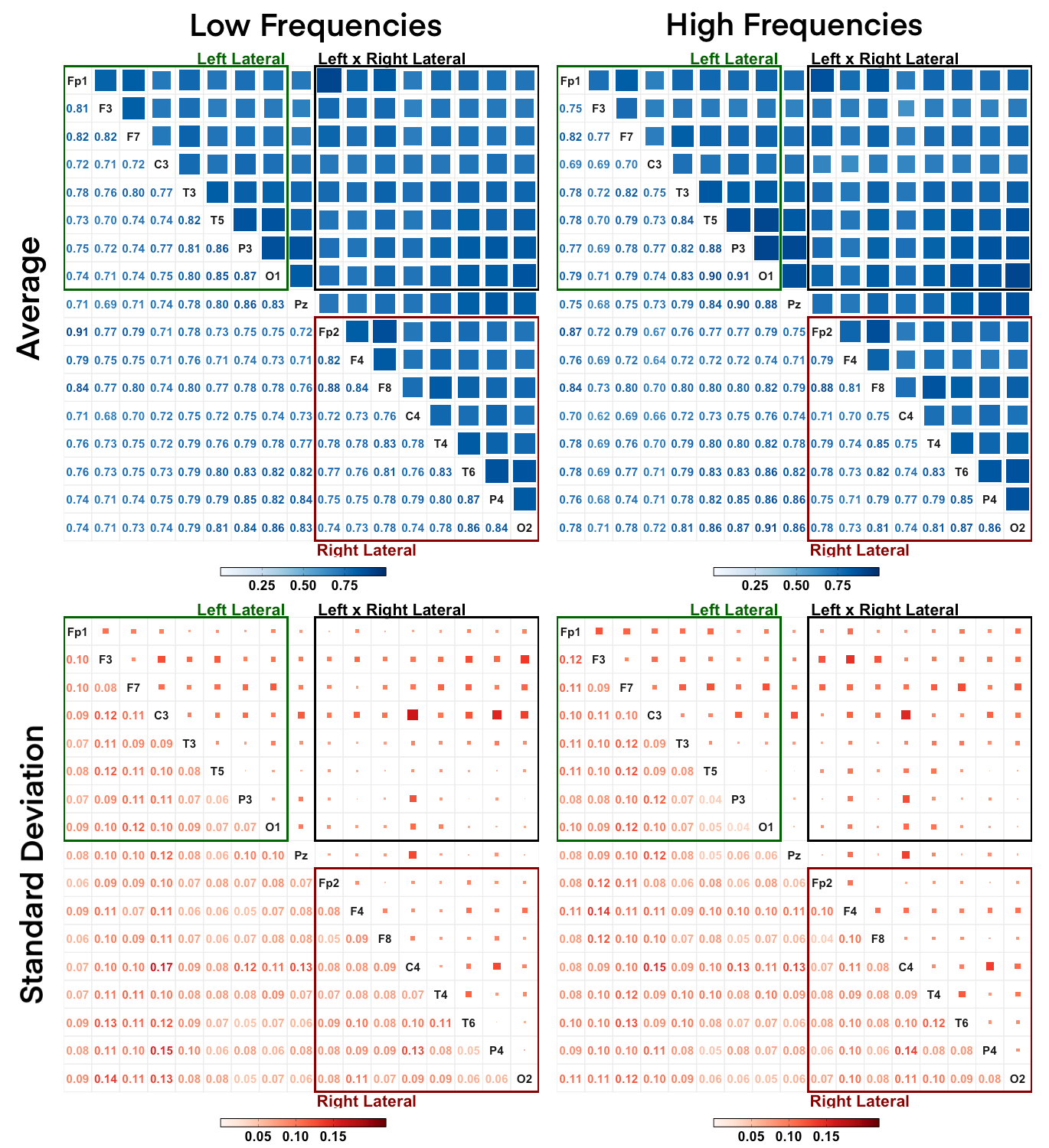}
\caption{Average (top) and standard deviation (bottom) of spectral connectivity persistence matrices for low (left) and high (right) frequencies, calculated using the HCC method, among \textbf{non-seizure} patients. In each plot matrix, the numbers (lower triangle) are the actual values, the shaded squares (upper triangle) visualize the values where larger and darker shade represent larger magnitude, and the boxes highlight the channels in the left lateral (upper left) and right lateral (lower right) regions, and the cross-lateral interactions (upper right).}
\label{fig:permat_hcc_ns}
\end{figure}

\begin{figure}[!th]
\centering
\includegraphics[width=0.9\textwidth]{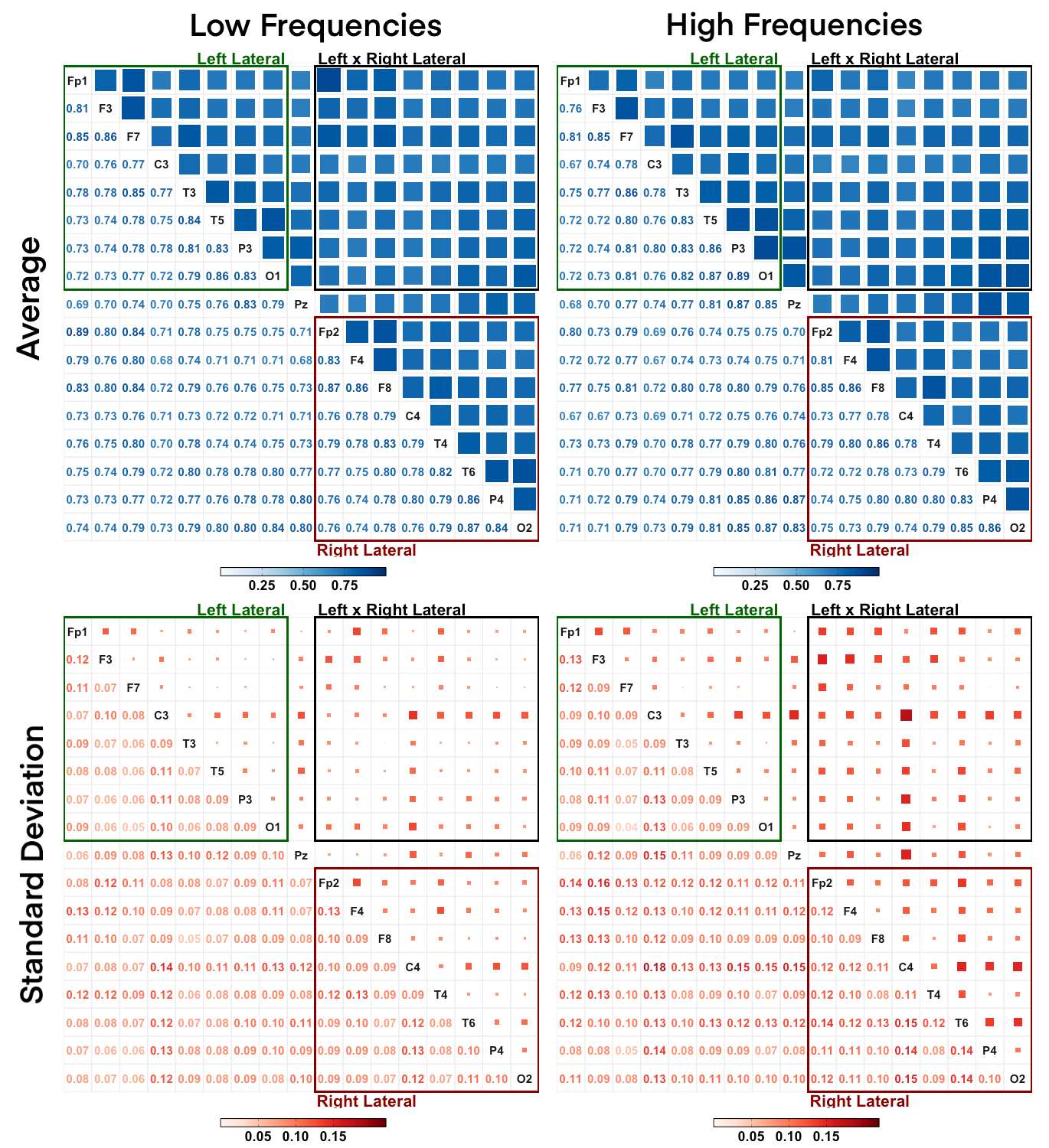}
\caption{Average (top) and standard deviation (bottom) of spectral connectivity persistence matrices for low (left) and high (right) frequencies, calculated using the HCC method, among \textbf{seizure} patients. In each plot matrix, the numbers (lower triangle) are the actual values, the shaded squares (upper triangle) visualize the values where larger and darker shade represent larger magnitude, and the boxes highlight the channels in the left lateral (upper left) and right lateral (lower right) regions, and the cross-lateral interactions (upper right).}
\label{fig:permat_hcc_s}
\end{figure}

Figures~\ref{fig:permat_hcc_ns}~and~\ref{fig:permat_hcc_s} summarize the average persistence of channel clustering and its variability across non-seizure and seizure patients, respectively, as estimated using the HCC method, for both low and high frequencies. The HCC-based spectral connectivity persistence matrices exhibit uniformly high co-clustering proportions (all \(\ge 62\%\)), making it difficult to
distinguish adjacent from non-adjacent channels. There is also no clear difference between the persistence of cluster membership among the two patient groups (neither from the average persistence nor from the consistency of channel clustering). Moreover, the variability in connectivity persistence is not reflective of seizure occurrences. Specifically, we only observe large variations in cluster membership related to the central channels C3 and C4, which may not be related to the seizure experienced by the neonatal patients. This is one drawback
 of the HCC method. That is, it unable to provide intuitive characteristics about seizure occurrences. This may be due to the nature of the method itself as it focuses on the entire range of values of the signal amplitudes. Hence, it may not be suitable for applications that deals with extreme behaviors such as seizures, which, on the other hand, are appropriately modeled/described by the Club Exco method. Thus, our novel Club Exco method, which can serve both as an exploratory and a confirmatory tool, offers a new approach to better understand brain dynamics in the presence of abnormalities.

\section{Conclusions\label{sec:conclusion}}
This paper introduces Club Exco, a novel clustering method grounded in extreme value theory, designed to identify brain communities characterized by co-occurring extreme EEG amplitudes. By focusing on tail behavior rather than average signal characteristics, Club Exco provides a distinctive perspective on brain connectivity—particularly relevant in pathological states marked by extreme signal fluctuations, such as seizures. When applied to a neonatal EEG dataset comprising 17 patients with clinically confirmed seizures and 13 patients with no seizures during approximately one hour of continuous monitoring, our method revealed meaningful differences in connectivity patterns between the two groups. These findings offer novel insights into seizure localization and propagation. We also compared Club Exco to an alternative clustering approach, HCC. The results indicated that Club Exco was able to uncover seizure-associated connectivity patterns—especially in both low and high frequency bands—that were not captured by the HCC method.

Our method enabled us to identify particular phenomena in certain patients where persistent clustering patterns, as described by our ECP matrices, have initial propagation nodes located outside the regions where seizures where diagnosed. A comprehensive analysis is required in a larger dataset with patients with focal seizures to validate this observation.

In addition, our approach allowed us to detect synchronization patterns of extreme values that may not be entirely explained by common connectivity or network models. However, further analysis is required to explore the potential of Club Exco as a quantitative biomarker to predict treatment responses in particular frequencies \citep{IctalConnectivityChildhood-Tenney-2018} or broadband \citep{StrengthStabilityEEG-Shrey-2018}, improve identification of seizure onset zones \citep{ConnectivityHighFreq-PintoOrellana-2024} where other non-extreme connectivity metrics have shown promising results.

Compared to traditional coherence-based clustering approaches, Club Exco provides more targeted information about rare, high-amplitude neural events. These findings highlight the value of incorporating extremal dependence structures into brain network analysis and open avenues for future research in neurological disorder diagnostics and monitoring, particularly in nonstationary and high-dimensional EEG settings.

Several promising avenues remain for extending the Club Exco framework toward clinical utility. One important direction is the development of diagnostic metrics derived from the ECP matrices. Specifically, we propose to investigate whether global or regional summaries of ECP—such as mean connectivity or its deviation from a normative baseline—can be used to classify patients as seizure or non-seizure cases. Our preliminary findings suggest that non-seizure patients exhibit lower variability in ECP values, which could serve as a stable reference profile. Building on this, future work could explore z-score normalization using non-seizure patients as a control group, enabling per-patient standardization of seizure data. Such a framework may highlight individual deviations in brain-wide extremal connectivity and facilitate patient-level phenotyping.

Furthermore, while the current dataset does not include detailed annotations regarding seizure type (e.g., focal vs. generalized), future applications of Club Exco to datasets with richer clinical metadata could evaluate whether distinct ECP patterns are associated with seizure subtypes. This would open the door to using extremal brain network characteristics not only for seizure detection but also for classification and treatment monitoring. Finally, we suggest quantifying the relationship between ECP magnitude and seizure burden—for example, correlating ECP scores with the proportion of time spent in ictal states during recordings. If a strong association is established, ECP-derived metrics could serve as quantitative biomarkers for seizure severity and progression, providing valuable support for clinical decision-making in both diagnostic and post-treatment contexts.

\bibliographystyle{plainnat}
\bibliography{ref}

\end{document}